\pdfoutput=1
\documentclass[fleqn,usenatbib,usedcolumn]{mnras}
\usepackage{amsmath}
\usepackage[british]{babel}
\usepackage{txfonts}
\let\la=\lesssim

\usepackage{graphicx}

\usepackage{amssymb}
\usepackage[T1]{fontenc}
\hypersetup{pdfauthor={R. Hirai},
               pdftitle={Formation Scenario of the Progenitor of iPTF13bvn Revisited},
               pdfkeywords={stars: massive -- binaries: close -- supernovae: individual: iPTF13bvn},
               bookmarksnumbered=true}
\newcommand{\msun}{\textnormal{M}_\odot}
\newcommand{\rsun}{\textnormal{R}_\odot}

\title[Progenitor of iPTF13bvn Revisited]{Formation Scenario of the Progenitor of iPTF13bvn Revisited}
\author[R. Hirai]{
Ryosuke Hirai$^{1}$\thanks{E-mail: hirai@heap.phys.waseda.ac.jp}
\\
$^{1}$Advanced Research Institute for Science and Engineering, Waseda University, 3-4-1, Okubo, Shinjuku, Tokyo 169-8555, Japan\\
}

\date{Accepted XXX. Received YYY; in original form ZZZ}

\pubyear{2016}

\begin{document}
\label{firstpage}
\pagerange{\pageref{firstpage}--\pageref{lastpage}}
\maketitle

\begin{abstract}
The formation scenario of the progenitor of iPTF13bvn has been revisited. iPTF13bvn is unique in the sense that a corresponding pre-supernova image has been identified. This has enabled us to strongly constrain the nature of its progenitor. From the pre-supernova image, light curve and the latest observations, it is currently widely accepted that the progenitor of iPTF13bvn was in a binary system. The fast decline in the light curve suggests a progenitor mass of $\sim3.5\msun$, and the upper limit on the remaining companion is $\sim20\msun$. Recent works suggest that binary evolution models involving common envelope episodes can satisfy all the observational constraints. We have examined the common envelope scenario based on latest knowledges on common envelope evolution. We have found that the common envelope scenario for the progenitor of iPTF13bvn seems not to be suitable since it can not explain the pre-supernova radius. We also propose an alternative model with a black hole companion. Stellar evolution calculations with a large black hole companion were carried out and succeeded in satisfying all observational constraints. However, more studies should be carried out to explain the origin of the large black hole companion.

\end{abstract}


\section{Introduction}
iPTF13bvn is the only type Ib supernova (SN) so far known to have a corresponding pre-explosion image. Ever since the first detection by \citet{cao13}, this pre-supernova (SN) image combined with information from the light curve has helped us deeply constrain the properties of the progenitor star. Hydrodynamical modelling and analytical fits to the light curve show that the ejecta mass was small, corresponding to a progenitor mass of $\sim3$--$4\msun$ \citep[]{ber14,fre14,sri14}. Such small mass progenitors are difficult to produce with single star evolution models, so it is now widely accepted that the progenitor had undergone binary interactions \citep[]{ber14,fre14,sri14,eld15,eld16}. 

There are two main channels of binary interactions; Roche lobe overflow and common envelope (CE) phases. Roche lobe overflow is a form of stable mass transfer, in which a Roche lobe filling star spills some of its mass from the outer envelope to the companion star through the inner Lagrangian point. It is not clear how much of the transferred mass will be retained by the accretor, but the secondary will become massive in general. On the other hand, CE phases are the consequence of unstable mass transfer, where the companion star plunges into the envelope of the evolved primary. If there is enough energy in the system to eject the entire envelope of the primary, it will end up as a close binary consisted of the core of the primary orbiting its companion. If not, the plunged-in star will fall to the centre, and can be regarded as a stellar merger. This process was first introduced to explain the formation of X-ray binaries \citep{van76}, but it is now commonly used to explain the formation of many close binaries with compact object components \citep[]{bel02,kal07}.

Earlier works on iPTF13bvn have favoured the stable mass transfer scenario for the formation of the progenitor \citep[]{ber14}. The initial mass ratio should have been close to unity for the mass transfer to be stable, and in the end the secondary will be much larger as a consequence of the mass accretion. The final companion mass predicted in this scenario was $18\lesssim m_2/\msun\lesssim40$, which will be bright enough so that we will be able to detect it three years after the explosion. But it turned out that the expected companion did not show up, ruling out this scenario \citep[]{fol16}. After this observation, the CE scenario has now become the current favourite. \citet{eld16} has showed some evolutionary models with CE phases which can produce the compact progenitor which is consistent with the pre-explosion image, with a fairly small mass companion. However, the whole process and outcome of CE evolution is still poorly understood, and we should be careful about how we treat CE phases in calculations \citep[][and references therein]{iva13}.

In this paper we will revisit the observational constraints on the progenitor of iPTF13bvn, and check the relevance of the CE scenario. We find that the CE scenario has a difficulty in explaining the final radius of the progenitor, which may be critical. We then propose another possible scenario which involves stable mass transfer with a black hole (BH) companion, and show some evolutionary tracks that are consistent with observation. This paper is structured as follows. We will first review the observational constraints on the progenitor of iPTF13bvn in section 2. In section 3 we will reconstrain the progenitor's position on the Hertzsprung-Russel (HR) diagram using the observational data. We will then discuss the relevance of the CE scenario in section 4 and suggest an alternative scenario in section 5. We will summarize and conclude our results in section 6.

\section{Summary of Observational data}\label{sec:review}
The rich observational data for the SN iPTF13bvn has enabled us to place strong constraints on the properties of the progenitor. In this section we will review and summarize the observational data and the analyses made in previous works.

\subsection{Host Galaxy Properties}
iPTF13bvn was first discovered by the intermediate Palomar Transient Factory in June 2013, in the galaxy NGC 5806. There is a wide scatter in the estimated host galaxy properties among various groups. For the extinction, \citet{cao13} suggest a host galaxy colour excess of $E(B-V)${\small $_\textrm{host}$}$=0.0278$ mag using Na \textsc{i} D absorption lines from their high resolution spectroscopy data. \citet{ber14} derived a higher reddening value of $E(B-V)${\small $_\textrm{host}$}$=0.17\pm0.03$ mag, assuming an intrinsic colour law based on observational samples by the Carnegie Supernova Project. This was supported by \citet{sri14} from a different analysis. \citet{ber14} also measured the Na \textsc{i} D lines and obtained $E(B-V)${\small $_\textrm{host}$}$=0.07$ or 0.22 depending on the model used. An intermediate value $E(B-V)${\small $_\textrm{host}$}$=0.08^{+0.07}_{-0.04}$ mag was suggested by \citet{fre16}, based on an assumption that the intrinsic colour of iPTF13bvn was similar to that of SN2011dh. This is consistent with all other values within the uncertainties. 

The distance to the galaxy also holds a large uncertainty. Many works in the literature use $22.5_{-3.4}^{+4.0}$ Mpc, $\mu=31.76\pm0.36$ for the distance and distance modulus which are taken from \citet{tul09}. More recent works use the updated distance of $26.8^{+2.6}_{-2.4}$ Mpc, $\mu=32.14\pm0.20$ by \citet{tul13}, or the mean value of all estimates $25.8\pm2.3$ Mpc, $\mu=32.05\pm0.20$ provided by the NASA/IPAC Extragalactic Database (NED). In this paper, we adopt $E(B-V)${\small $_\textrm{host}$}$=0.08^{+0.07}_{-0.04}$ mag for the extinction value, and $25.8\pm2.3$ Mpc for the distance to the host galaxy.

\subsection{Pre-Explosion Image}
\begin{figure}
 \includegraphics[width=\linewidth]{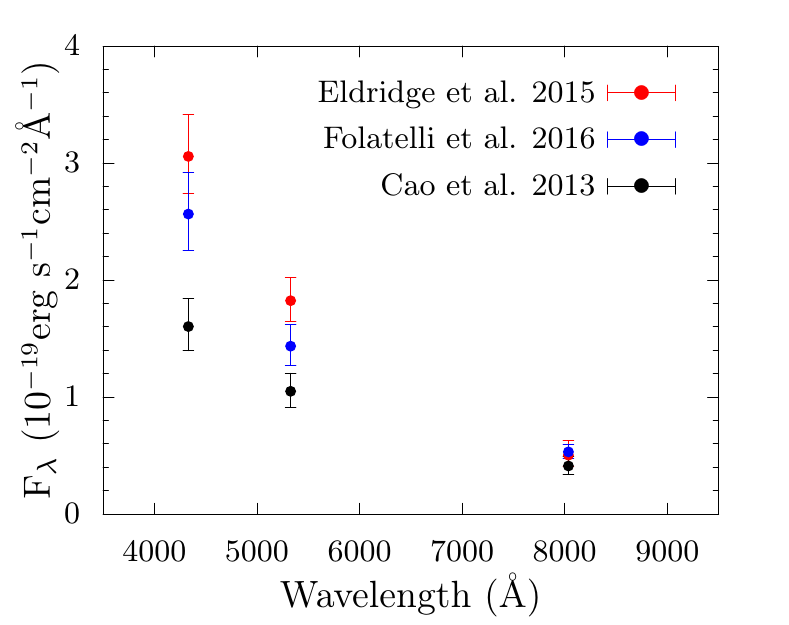}
 \caption{Observed pre-SN flux of the location of iPTF13bvn. Each colour shows the reported flux by Eldridge et al. 2015 (red), Folatelli et al. 2016 (blue) and Cao et al. 2013(black).\label{preSNflux}}
\end{figure}

It was first reported by \citet[]{cao13} that they have identified a progenitor candidate at the location of the SN from an observation made by the HST in 2005. Based on their results, various studies were carried out to construct a progenitor model consistent with this pre-SN source and the light curve of the SN itself. Early works showed that they were consistent with a Wolf-Rayet star progenitor \citep[]{cao13,gro13}, but binary progenitors were also suggested later on \citep[]{ber14,fre14,sri14}. \citet{eld15} re-analysed the HST data of the progenitor candidate, and found that the reported magnitude by \citet[]{cao13} was lower than that of their analysis by $\sim0.7$ mag. Their new magnitude was supported by other following studies \citep[]{fol16}, although they seem to have misread the results by \citet{eld15}. In Fig. \ref{preSNflux} we compare the fluxes calculated from the reported magnitudes by \citet{cao13}, \citet{eld15} and \citet{fol16}. 
It can be seen that the latter two have some overlaps within the uncertainties, but the three results are rather inconsistent with each other overall.
 It is not clear why there is such a large discrepancy between the different analyses. It is suggested that the differences of the parameters used in the data reduction may have amplified very small errors \citep[]{eld15}. The late-time view of the SN position may help us improve our knowledge of the pre-SN image by refining the background information \citep[]{mau14}.

\subsection{Light Curve}
Another constraint can be placed on the progenitor from the light curve of the SN itself. Although there were very high ejecta mass estimates ($M_\textrm{ej}\sim8\msun$) in the early works \citep[]{cao13,gro13}, the relatively fast decline in the light curve of iPTF13bvn showed that the ejected mass should have been small. According to hydrodynamical modelling, the ejecta mass was estimated to be $M_\textrm{ej}\approx2\msun$ which indicates that the progenitor was a $M_\textrm{He}\approx3.5\msun$ He star \citep[]{ber14,fre14}. A simple analytical fit also suggested $M_\textrm{ej}\sim1.5$--$2.2\msun$ \citep[]{sri14}. 

These analyses ruled out all single star evolution models. The minimum possible mass achieved by single star models with realistic wind mass-loss rates is $\sim8\msun$. The only other way to remove the entire hydrogen envelope up to the observed mass is by binary interactions (in our current understandings).

\subsection{Other Constraints}
There are some attempts to infer the zero-age main sequence (ZAMS) mass from late time spectra. The [O I]$\lambda\lambda6300, 6343$ emission lines can be used to estimate the mass of ejected oxygen \citep[]{jer15}. By fitting the late time spectrum with ejecta models, the ejected oxygen mass was estimated to be $\sim0.3\msun$. \citet{fre16} associated this mass with a star with ZAMS mass $\sim12\msun$ based on 1D single star nucleosynthesis calculations by \citet{woo07}. \citet{kun15} derived the ZAMS mass to be $\lesssim15$--$17\msun$ using nucleosynthesis models by \citet{nom97,lim03,rau02}. Both values have large uncertainties due to the complexity in modelling the star and explosive nucleosynthesis. For example, all models do not take into account the possible multidimensional effects such as turbulent mixing in the core, that may change the nucleosynthesis yields significantly \citep[]{smi14}. Therefore this constraint should be treated carefully when comparing with stellar models.

Latest observations by \citet{fol16} have revealed that the progenitor of iPTF13bvn has disappeared, and also placed an upper limit on the brightness of the possible companion. The magnitudes in each band were $m_\textrm{F225W}>26.4, m_\textrm{F435W/F438W}=26.62\pm0.14, m_\textrm{F555W}=26.72\pm0.08, m_\textrm{F814W}=26.03\pm0.15$ in June 2016. Especially the strict constraint in the F225W band ruled out most luminous companions as predicted in \citet{ber14}, and they stated that only late-O type stars with masses $\lesssim20\msun$ are possible assuming that it is not obscured by newly created dust. \citet{eld16} derived a slightly brighter magnitude from the same data, $m_\textrm{F438W}=26.48\pm0.08$ and $m_\textrm{F555W}=26.33\pm0.05$.

\section{Reconstraining the Progenitor System}\label{sec:method}

\subsection{Methodology}

Using these constraints, we attempt to pin down the position of the progenitor on the HR diagram. We use the reddening law of \citet{car89} for the extinction correction, with the standard coefficient $R_\textsc{v}=3.1$ and combining the reddening values from the host galaxy and the Milky way foreground \citep[$E(B-V)${\small $_\textsc{mw}$}$=0.0437$ mag;][]{sch11}. For each combination of luminosity and temperature $(L,T_\textrm{eff})$ and assuming that the star can be approximated as a black body\footnote{This is a good approximation for low-mass He star progenitors, since they do not have optically thick winds \citep[]{yoo12}. We also assume that the flux is dominated by the primary star in the bands considered here. If the binary companion is on the main sequence, it will be optically fainter than the cool envelope of the low-mass progenitor.}, we can calculate the flux in each band after applying the extinction correction and giving a distance. If there is a consistent combination of $E(B-V)$ and distance within their uncertainties where all three calculated fluxes fit in the error bars of the observation (see Fig.\ref{preSNflux}), we consider the combination ($L,T_\textrm{eff}$) is ``allowed''. This procedure is similar to the selection process of matching models in \citet{eld15,eld16}. However, our selection is more strict since we require to find a combination of distance and extinction value that is consistent for all three bands.

In the same way we can derive a ``forbidden'' region for the secondary star. We assume that the data obtained by \citet{fol16} are upper limits. Then for each combination of ($L,T_\textrm{eff}$), we calculate the flux in each of the four bands assuming that it is a black body. If the flux in any band exceeds the upper limit, we mark the combination as ``forbidden''.

\subsection{HR Diagram Constraints}\label{sec:results}

\begin{figure}
 \includegraphics[width=\linewidth]{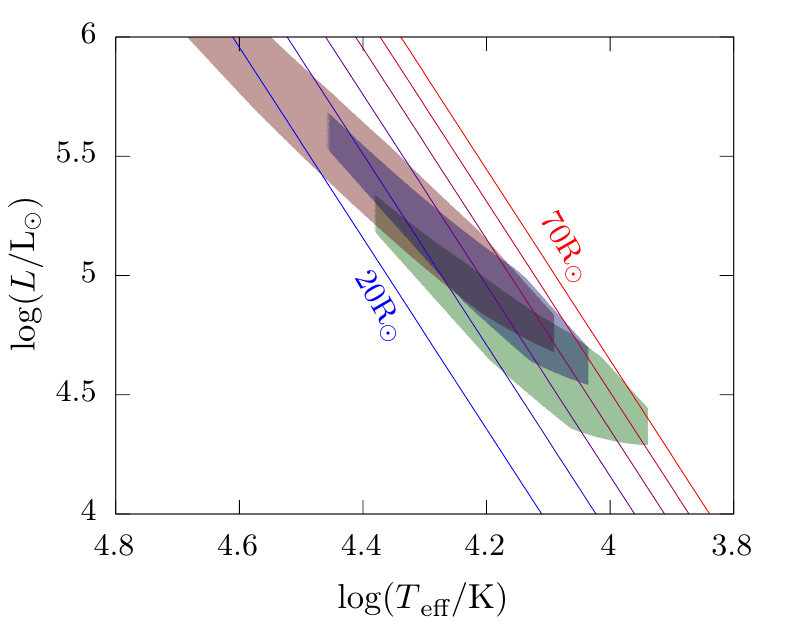}
 \caption{Allowed regions of the progenitor of iPTF13bvn on the HR diagram. Colours of the shaded regions show the results that fit the observed magnitudes obtained by \citet[][green]{cao13}, \citet[][red]{eld15} and \citet[][blue]{fol16}. Lines correspond to constant radii drawn with intervals of 10 R$_\odot$.\label{allowed}}
\end{figure}

\begin{figure}
 \includegraphics[width=\linewidth]{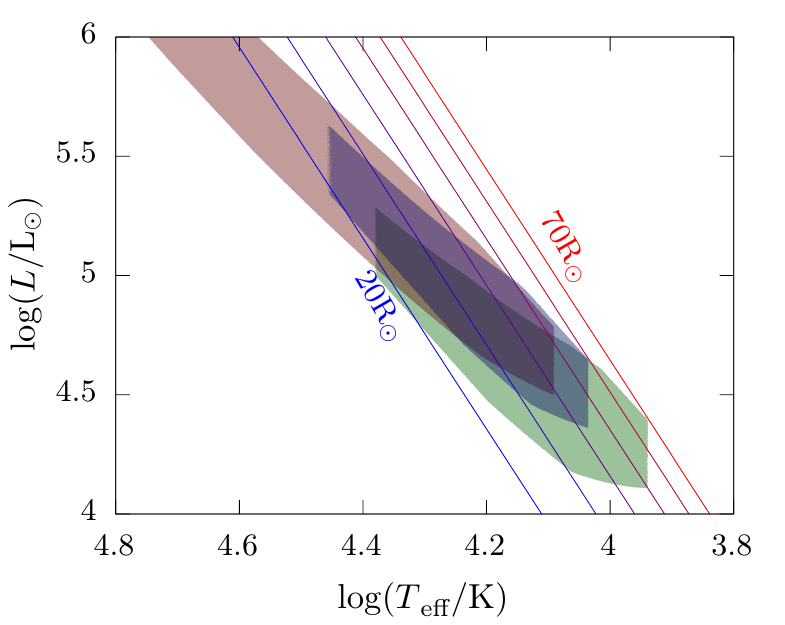}
 \caption{Same as Fig.\ref{allowed} but calculated using the smaller value ($22.5^{+4.0}_{-3.4}$Mpc) for the distance to the host galaxy.\label{allowed2}}
\end{figure}

Fig.\ref{allowed} shows the calculated allowed regions for the progenitor on the HR diagram, i.e. the progenitor for iPTF13bvn should have been positioned in the shaded region eight years before the explosion. The size and place of the allowed region strongly depends on which observational data are used. It also depends on the assumed host galaxy properties. For example, in Fig.\ref{allowed2} we show the same plot but using a smaller distance ($22.5^{+4.0}_{-3.4}$Mpc) to the host galaxy. Smaller luminosities become allowed obviously because of the closer distance assumed. If we take a larger extinction value, the shape of the region will extend to the upper left direction. It should also be noted that the overlapped region is not particularly favoured because the three reported fluxes are not independent observations, but different analyses performed on the same data. 

From this analysis only, we can place a stringent constraint on the radius of the progenitor. In Figs.\ref{allowed} and \ref{allowed2}, we have overplotted lines of constant radii. Most parts of the allowed region are within 20--70R$_\odot$. Since the progenitor mass is constrained to very low masses ($\sim3.5\msun$), the luminosity should not be so high. Therefore, the progenitor had most likely been in the lower right end of the allowed region. This implies that the radius was larger than $\sim30\textrm{R}_\odot$.

\begin{figure}
 \includegraphics[width=\linewidth]{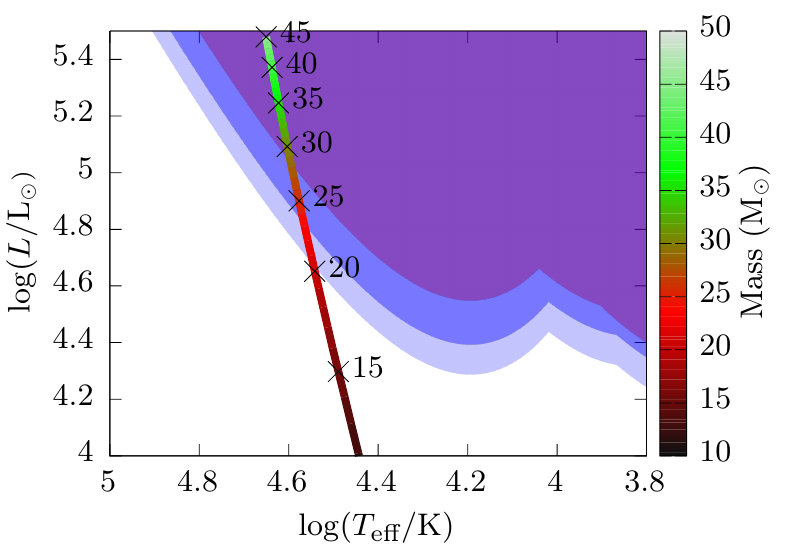}
 \caption{Forbidden regions of the possible companion star.  calculated from the post-explosion photometry combined with the fiducial parameters for the host galaxy (blue+purple), with a larger extinction value (purple) or with a larger distance (light blue+blue+purple). The line indicates the ZAMS stars coloured according to the mass.\label{forbidden}}
\end{figure}

The forbidden regions for the companion calculated from the post-explosion photometry are shown in Fig.\ref{forbidden}. With the fiducial set of parameters for the host galaxy ($E(B-V)${\small $_\textrm{host}$}$=0.08^{+0.07}_{-0.04}$ mag, $25.8\pm2.3$ Mpc), main sequence stars larger than $23\msun$ can be ruled out. A stricter constraint $m_2<20\msun$ can be placed if the host galaxy is closer ($22.5^{+4.0}_{-3.4}$Mpc), whereas the upper limit goes up to $m_2<29\msun$ if the larger extinction value $E(B-V)=0.17\pm0.03$ is true. It should be noted that these limits are rather overestimated. The line showed in Fig.\ref{forbidden} is the location of ZAMS stars, but the secondary will at least have an age equivalent to the lifetime of the primary. A star on the main sequence evolves slowly to the upper right in the HR diagram, so stars just outside the forbidden region will slide in eventually. Also, \citet{RH15} suggest that the SN ejecta can drive a shock into the companion star, injecting heat to the outskirts of the envelope. The heat excess will puff up the star to larger radii. This can lower the surface temperature temporarily, taking the star to the right on the HR diagram, which will also strengthen the upper constraint. Having these in mind, we consider that the upper limit $m_2\lesssim20\msun$ noted by \citet{fol16} is reasonable.

As we have seen in the previous section, the progenitor of iPTF13bvn was most likely a $\sim3.5\msun$ He star. Stars that have such a large He core must have had a ZAMS mass of $M_\textsc{zams}\gtrsim10\msun$. This means that the progenitor should have lost at least $\gtrsim7\msun$ of its matter on the course of its evolution. Single star models have been excluded already because strong stellar winds in the Wolf-Rayet phase is not enough to produce such small progenitors \citep[]{ber14,fre14,sri14}. This leads us to resort to binary evolution models. If the mass was stripped off via stable mass transfer, the companion star should have grown rather large \citep[$18\lesssim m_2/\msun\lesssim45$;][]{ber14}. However, such large companions have been ruled out. The other possible scenario to strip off such a large amount of mass is by experiencing a CE phase. In this way the primary can lose most of its hydrogen envelope, even with relatively small companions. The fact that a CE process is necessary was already suggested by the calculations in \citet{eld16}. From what we have shown in this section, all the observational facts seem to favour the CE scenario.

\section{Common Envelope Scenario}\label{sec:discussion}
We have shown that the progenitor of iPTF13bvn has most likely experienced a CE phase. In this section, we will first briefly review the current status on CE evolution in general. Then we will inspect the CE scenario for the progenitor of iPTF13bvn by modelling the post-CE structures of stars with various ZAMS masses, and checking whether their final position on the HR diagram lies within the allowed region. We use a different treatment for CE evolution from \citet{eld15,eld16}. Using those results, we also discuss the final separation which is strongly related with the CE efficiency and check the ejectability of the envelope. 

\subsection{Common Envelope Evolution}

The main focus of CE studies is whether or not the system can eject the envelope. In most population synthesis studies, the outcome is estimated by the ``energy formalism'' or ``alpha-formalism'', which is expressed as belows \citep[]{web84,ibe84}.
\begin{eqnarray}\label{eq:alphaformalism}
 E_\textrm{env}=\alpha_\textsc{ce}\left(-\frac{Gm_1m_2}{a_i}+\frac{Gm_{1,c}m_2}{a_f}\right)
\end{eqnarray}
$E_\textrm{env}$ is the binding energy of the envelope, $G$ is the gravitational constant, $m_1, m_2, m_{1,c}$ are the masses of the primary, secondary and core of the primary respectively, $a_i$ and $a_f$ are the initial and final separations respectively. It assumes that as the secondary star plunges into the envelope, the orbital energy is somehow transferred to the envelope to unbind it. The mass of the secondary is assumed to be unchanged before and after the CE phase, because the time-scale of the CE phase is much shorter than the thermal time-scale of the secondary, so there will be almost no accretion \citep[]{iva13,mac15}. $\alpha_\textsc{ce}$ is a parameter expressing the efficiency of the energy conversion. The value of $\alpha_\textsc{ce}$ should be calibrated somehow by observation or theory, but so far there is no guiding principle. Instead, many studies simply take $\alpha_\textsc{ce}=1$ or leave it as a free parameter to study the dependences of the resulting populations. The binding energy $E_\textrm{env}$ is often estimated by 
\begin{eqnarray}
 E_\textrm{env}=\frac{Gm_1m_\textrm{1,env}}{\lambda R_1}
\end{eqnarray}
where $m_\textrm{1,env}=m_1-m_{1,c}$ is the envelope mass, and $R_1$ is the radius of the primary. $\lambda$ is another parameter introduced to characterize the structure of the star \citep[]{dek90}. Although there are several studies deriving a fitting formula for the value of this parameter \citep[]{xu10,wan16}, many studies combine the uncertainties of the two parameters and simply take $\alpha_\textsc{ce}\lambda=1$ with no strong reasoning. Given the masses $m_1$ and $m_2$, an estimate for the core mass $m_{1,c}$, the initial separation and a value for the parameters, we can calculate the resulting separation $a_f$ of the binary. The criterion for a successful ejection is that both of the post-CE binary components do not overfill their Roche lobes. However, there are still issues regarding the radius of the post-CE remnant \citep[]{hal14}.

There are of course some other attempts to understand CE phases such as from observation and simulations. As the secondary plunges into the envelope, it is considered that a small amount of mass is ejected due to the shock created at the interface, and this can be observed as a ``luminous red nova''. But the typical ejecta mass is very small, making detections difficult due to the low luminosity. The situation has started to change in the past few years, and now there is a rapidly growing number of candidates for the detection of a CE onset \citep[]{iva13b,che14,mac15b,bla16}.  However, much more data are required to be able to constrain CE physics from observation. Hydrodynamical simulations have been performed to investigate the internal physics of a CE phase but the huge dynamical range ($\sim10^{13}$) makes it extremely computationally expensive. Several groups have already attempted large-scale simulations, but it is still hard to extract general features from the small number of models studied \citep[]{ric12,pas12,nan14,iac16,ohl16a,ohl16b}.

\subsection{Post-CE Structure}\label{postCE}
\begin{figure}
 \includegraphics[width=\linewidth]{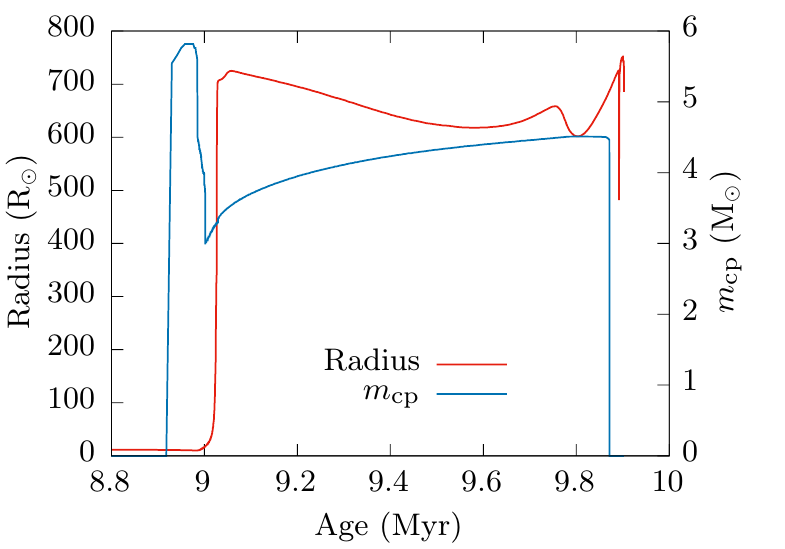}
 \caption{Time evolution of the radius and $m_\textrm{cp}$ for a 17$\msun$ star with metallicity $Z=0.02$.\label{r_mcp}}
\end{figure}

The progenitor of iPTF13bvn should have a temperature and luminosity in the allowed region shown in Fig.\ref{allowed}, eight years before the explosion. To see what kinds of stars can end up in this region, we carry out stellar evolution calculations to model the pre-SN state of stars which have experienced CE evolution. All calculations were carried out using the stellar evolution code \texttt{MESA} \citep[version 8645;][]{MESA1,MESA2,MESA3}. For convection, we use the mixing length theory, with the Ledoux criterion and a mixing length parameter 1.6. We use the prescription by \citet{dej88} for the wind mass-loss rate. To create post-CE stellar structures, we follow the procedures taken in \citet{iva11}. First, we evolve a star until it enters the hydrogen shell burning phase. Once the stellar radius expands up to a certain value, at which we assume the CE phase kicks in, we search for the mass coordinate of the ``maximum compression point'' in the hydrogen burning shell $m_\textrm{cp}$. This is currently assumed to be the best estimate for the bifurcation point of the core and envelope \citep{iva11,iva13}. After that  we give an extremely high mass-loss rate of $0.1\msun \textrm{yr}^{-1}$ artificially\footnote{This corresponds to a CE timescale of $\sim100$ yr, which is the typical CE timescale.}, and wait until the mass drops to $m_\textrm{cp}$. Once the mass has dropped to $m_\textrm{cp}$, we switch off the artificial mass-loss and evolve the star until it starts burning neon at the centre. A star burning neon will explode within a few more days. The radius at which we start the artificial mass loss is not so important since the time-scale of the expansion is smaller than the time-scale of the core mass growth. This can be checked in Fig.\ref{r_mcp} where we show an example of the evolution of the radius and the core mass in the late stages. The core mass increases by only $\sim1\%$ during the expansion. Fig.\ref{r_mcp} is for a 17$\msun$ star, but the same applies to all stars in the mass range we used. 

\begin{figure}
 \includegraphics[width=\linewidth]{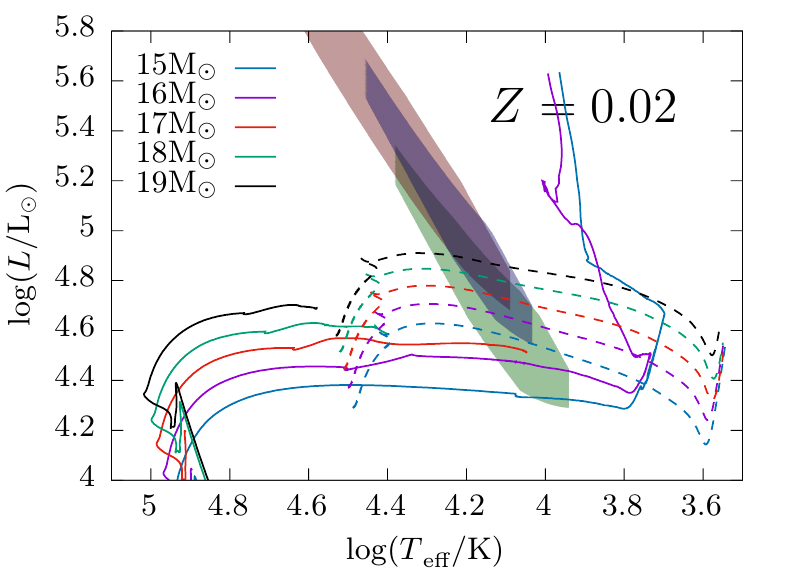}
 \caption{Evolutionary tracks of stars with a metallicity $Z=0.02$ on the HR diagram. Dashed lines are for before the CE phase, and solid lines are for after the CE phase up to eight years before collapse. The shaded regions are taken from Fig.\ref{allowed}.\label{evotrackZ0.02}}
\end{figure}

\begin{figure}
 \includegraphics[width=\linewidth]{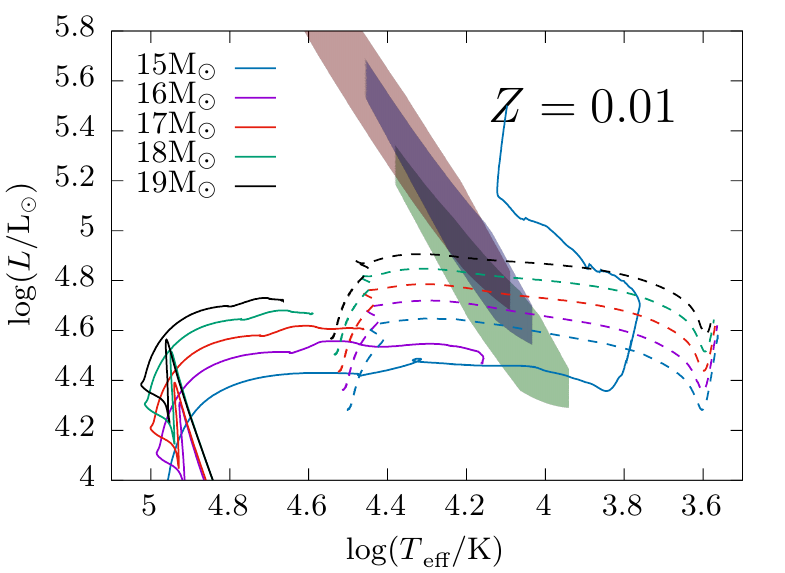}
 \caption{Same as Fig.\ref{evotrackZ0.02} but with a metallicity $Z=0.01$.\label{evotrackZ0.01}}
\end{figure}

\begin{figure}
 \includegraphics[width=\linewidth]{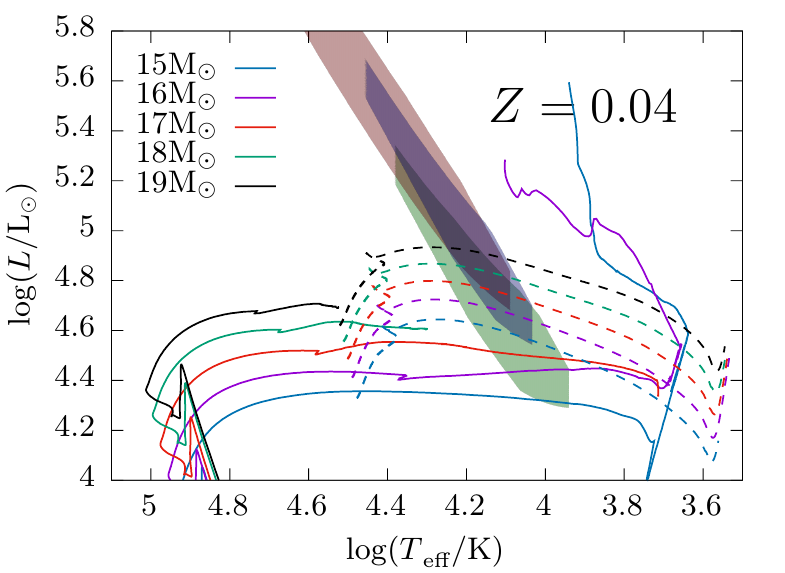}
 \caption{Same as Fig.\ref{evotrackZ0.02} but with a metallicity $Z=0.04$.\label{evotrackZ0.04}}
\end{figure}

In Fig.\ref{evotrackZ0.02} we show the evolutionary tracks of our post-CE stars with an initial metallicity $Z=0.02$. All stars follow similar tracks from ZAMS to the red giant phase (dashed line). After that we switch on the artificial mass-loss, and at the end of the CE phase all stars end up on the left end of the HR diagram. Then the stars evolve towards core-collapse. The lighter stars ($M_\textsc{zams}=$15--16$\msun$) evolve from left to right, crossing over the allowed region and then follow very complex paths. This complex evolution is probably not real, so we simply stop our calculation after the track has moved away significantly. The heavier stars ($M_\textsc{zams}=$17--19$\msun$) also evolve with constant luminosity from left to right, but starts to collapse somewhere on the way towards the allowed region. We only plot up to eight years before collapse, since the pre-SN image for iPTF13bvn was taken eight years before explosion. Only the $17\msun$ model ended up in the allowed region in our mass range. However, the final temperature -- or radius -- is very sensitive to the details of the calculation such as the mixing length or overshoot parameters or metallicity, so we can not derive a concrete conclusion about the best mass range. For example in Figs.\ref{evotrackZ0.01} and \ref{evotrackZ0.04} we show the evolutionary tracks for our lower and higher metallicity models. With lower metallicity the expansion of the stars are smaller than in Fig.\ref{evotrackZ0.02}, and somewhere between 15 and 16$\msun$ seems to be the matching model. Higher metallicity led to larger expansion, and the mass of the matching models increases.

 The ZAMS masses of our matching models are rather heavier than the matching models in \citet{eld16}. This may be due to the different treatments of the CE evolution. In their \texttt{BPASS} code, they use the usual RLOF rate but limit it by $\dot{M}=10^{-3}\msun$ for the mass-loss rate during CE evolution \citep[]{eld08}, and terminate when both stars reside within their Roche lobes. Their choice for the upper limit value is due to numerical reasons, and not motivated physically. A CE phase is a highly dynamical process, and the usual mass-loss rates that were derived assuming nuclear time-scale processes do not describe the dynamical nature of CE evolution well. The lower mass-loss rate will lead to a longer time-scale for the CE phase, giving more time for the core to grow. Together with their different termination criterion, their method will always leave a larger remnant than ours, which may possibly explain the discrepancy of the results.

It should also be noticed that the ZAMS masses of our matching models are within the range estimated from the nebular phase oxygen lines. But because in the CE scenario we remove the envelope before the core grows to its full size, the final ejected oxygen may be smaller than what we would expect from a progenitor with that ZAMS mass.

Although we have a matching model, the final temperatures in the stellar evolution calculations are not so reliable, so we will not conclude which models are the best. On the other hand, the luminosity is almost constant in the final stages, which is strongly correlated with the core mass. From the lower limit of the luminosity of the allowed region, we can place a rough lower limit $\sim2.5\msun$ on the core mass of the progenitor.

\subsection{Pre-CE Separation}

Here we will discuss the upper limit to the initial orbital separation of the progenitor system in the context of the CE scenario. There are two pathways known so far to initiate a CE phase. The first is via unstable mass transfer. Once the primary star fills its Roche lobe, a part of the outer envelope of the star will be transferred to the secondary through the inner Lagrangian point. This is the usual Roche lobe overflow. If the mass transfer is unstable, the star will eventually overfill the second Lagrangian point (only the primary component). Then a part of the envelope material will start trickling away from the system. This flow will take away angular momentum, shrinking the orbit even more, leading to a CE phase. The onset of an unstable mass transfer is usually computed by comparing the volume of the star with the primary component of the volume enclosed within the equipotential surface passing through the second Lagrangian point. The effective radius of this volume $R_{\textrm{L}_2}$can be approximated by
\begin{eqnarray}
 \frac{R_{\textrm{L}_2}}{a}&\approx&\frac{0.49q^{2/3}+0.27q-0.12q^{4/3}}{0.6q^{2/3}+\ln(1+q^{1/3})},\;\;\;  q\leqq1\\
&\approx& \frac{0.49q^{2/3}+0.15}{0.6q^{2/3}+\ln(1+q^{1/3})},\;\;\;  q\geqq1
\label{eq:L2}
\end{eqnarray}
where $a$ is the binary separation, $q\equiv {m_1}/{m_2}$ and $m_1, m_2$ are the primary and secondary masses \citep[]{egg11}. Thus the criterion for unstable mass transfer will be 
\begin{eqnarray}
 R>R_{\textrm{L}_2}
\end{eqnarray}
where $R$ is the radius of the primary.
The other path is via Darwin instability \citep{hut80,lai93}. This occurs when the tidal forces extract orbital angular momentum to spin up the stars, but there is not enough angular momentum left in the orbit to do so. The condition for this instability is 
\begin{eqnarray}
 J_\textrm{spin}>\frac{1}{3}J_\textrm{orb}
\end{eqnarray}
where $J_\textrm{spin}$ is the moment of inertia of the primary star and $J_\textrm{orb}$ is the moment of inertia of the orbit. $J_\textrm{orb}$ can be expressed as $J_\textrm{orb}=\mu a^2$ where $\mu=m_1m_2/(m_1+m_2)$ is the reduced mass.

\begin{figure}
 \includegraphics[width=\linewidth]{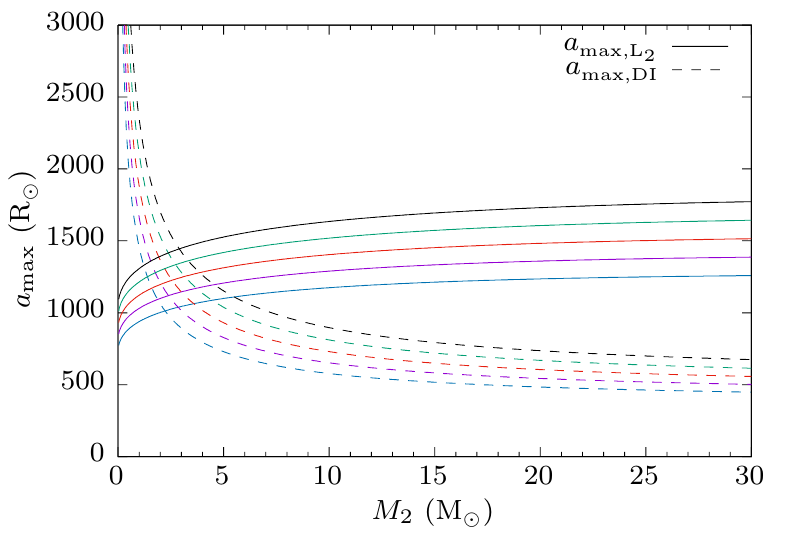}
 \caption{Maximum orbital separation as a function of the secondary mass $M_2$. Line colours express the primary mass, with the same colours as in Fig.\ref{evotrackZ0.02}\label{a_max}}
\end{figure}

In either of the cases, the CE phase will be initiated at the time when the primary star evolves to a red giant, and is rapidly expanding in size. Both the radius and moment of inertia of the star grow rapidly at this stage, and will eventually satisfy one of the criteria above, depending on the secondary mass. If $m_2$ is relatively large, $J_\textrm{orb}$ will be large, so the system is unlikely to be Darwin unstable and thus enters the CE phase via unstable mass transfer. The maximum possible separation for unstable mass transfer to occur can be estimated by 
\begin{eqnarray}
 a_{\textrm{max,L}_2}\approx R_\textrm{max}\frac{0.6q^{2/3}+\ln(1+q^{1/3})}{0.49q^{2/3}+0.15}
\end{eqnarray}
where $R_\textrm{max}$ is the maximum radius achieved in single star evolution. If $m_2$ is relatively small, $J_\textrm{orb}$ is small and the system will be Darwin unstable before the primary overflows the L$_2$ point. The maximum possible separation to be Darwin unstable $a_\textrm{max,DI}$ can be estimated by 
\begin{eqnarray}
 a_\textrm{max,DI}\approx \sqrt{\frac{3J_\textrm{spin,max}}{\mu}}
\end{eqnarray}
where $J_\textrm{spin,max}$ is the maximum moment of inertia obtained in the single evolution models. In Fig.\ref{a_max} we show the maximum possible orbital separation as a function of secondary mass. We have used $J_\textrm{spin,max}$ and $R_\textrm{max}$ obtained from single star evolution calculations with metallicity $Z=0.02$. The maximum separation is $\sim1000$--$1800\rsun$ throughout most of the mass range, which is determined by the L$_2$ overflow criterion. Larger separations would be possible only if the secondary mass was smaller than $\sim4\msun$.

\subsection{CE Efficiency parameter}\label{sec:alpha}
We will now constrain the $\alpha_\textsc{ce}$ parameter in this system to discuss the ejectability of the envelope. In usual population synthesis calculations, $\alpha_\textsc{ce}$ is given by hand, to calculate the final separation $a_f$. We will go the other way round, and use the constraints on $a_f$ to calculate a lower limit to $\alpha_\textsc{ce}$. Eq.\ref{eq:alphaformalism} can be rewritten as
\begin{eqnarray}\label{eq:minalpha}
 \alpha_\textsc{ce}&=&E_\textrm{env}\left(-\frac{Gm_1m_2}{a_i}+\frac{Gm_\textrm{cp}m_2}{a_f}\right)^{-1}\nonumber\\
&\geq&\frac{E_\textrm{env}a_f}{Gm_2m_\textrm{cp}}
\end{eqnarray}
The inequality can almost be regarded as an equality because the initial separation is usually much larger than the final separation, and thus the first term in the parenthesis can be ignored. We have a rough estimate on $m_\textrm{cp}$ from the observed ejecta mass. Therefore the important values that determines $\alpha_\textsc{ce}$ are $E_\textrm{env}$ and $a_f$.

The binding energy of the envelope is usually estimated by
\begin{eqnarray}\label{eq:Eenv1}
 E_\textrm{env}=-\int_{m_\textrm{cp}}^{m_1}\left(-\frac{Gm}{r}+\epsilon\right)dm
\end{eqnarray}
where $m_1$ is the total mass of the star and $\epsilon$ is the specific internal energy. But in order to take into account the relaxation of the core after the mass ejection, it should be calculated by comparing the total binding energies of the star before and after the CE event \citep[]{ge10}.
\begin{align}\label{eq:Eenv2}
 E_\textrm{env}&=E_{\textrm{bind},i}-E_{\textrm{bind},f}\nonumber\\
&=-\int_0^{m_{1,i}}\left(-\frac{Gm}{r}+\epsilon\right)dm+\int_0^{m_{1,f}}\left(-\frac{Gm}{r}+\epsilon\right)dm
\end{align}
Here the integration is taken over the whole star before (first term) and after (second term) the CE phase. For the model CE calculations in section \ref{postCE}, the values calculated by Eq.\ref{eq:Eenv1} overestimated the binding energy by $\sim10\%$. We use the values calculated by Eq.\ref{eq:Eenv2} in our following discussions.

The final separation is quite uncertain. The closest possible separation is when the post-CE primary star (and of course the secondary) does not overfill its Roche lobe. Using the post-CE radius obtained in the model CE simulations, we can calculate the minimum possible separation by assuming that one of the binary components exactly fills its Roche lobe. This can be expressed as 
\begin{eqnarray}
 a_{f,\textrm{min}}=\max\left(\frac{R_f}{f(q)},\frac{R_2}{f(q^{-1})}\right)
\end{eqnarray}
where $R_f$ is the post-CE radius of the primary star, $R_2$ is the secondary radius, and $q=m_\textrm{cp}/m_2$. $f(q)$ is a function fitted to the approximate Roche lobe radius \citep[]{egg83}.
\begin{eqnarray}
 f(q)\equiv\frac{0.49q^{2/3}}{0.6q^{2/3}+\log(1+q^{1/3})}
\end{eqnarray}

\begin{figure}
 \includegraphics[width=\linewidth]{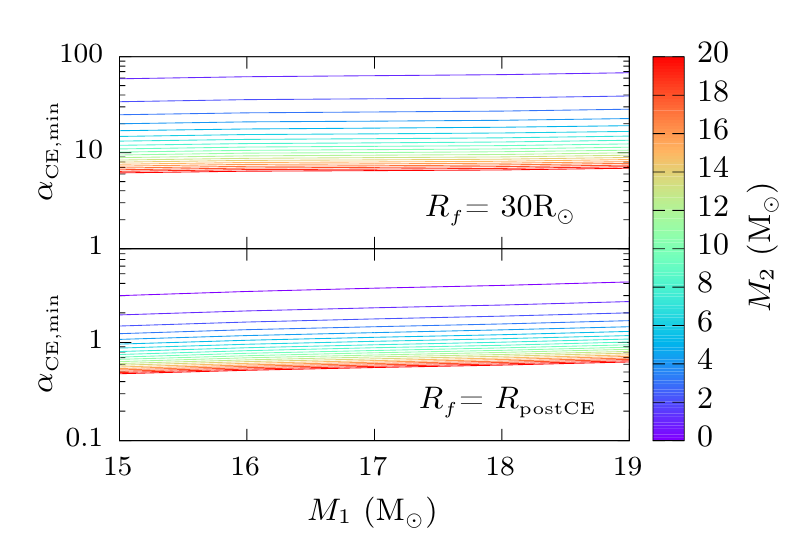}
 \caption{Lower limits on the $\alpha_\textsc{ce}$ parameter based on the assumption that the secondary never interacted with the primary again (upper panel), or the secondary has already been lost (lower panel).\label{fig:minalpha}}
\end{figure}

We can obtain the lower limit to $\alpha_\textsc{ce}$ as a function of the secondary mass by plugging in $E_\textrm{env}$ and $a_{f,\textrm{min}}$ into Eq.\ref{eq:minalpha}. This is shown in the lower panel of Fig.\ref{fig:minalpha}. We have used the ZAMS radius for $R_2$. The minimum value in our calculations was $\sim0.5$, which means that at least half of the orbital energy should be used to eject the envelope. The limit increases as the secondary mass decreases because of the decrease of the energy reservoir in the orbit. If the secondary was $\lesssim6\msun$, the lower limit exceeds unity, which suggests the presence of an extra energy source to eject the envelope. All models were limited by the secondary star radius filling its Roche lobe. If we consider a compact object as the companion, the final separation will be limited by the post-CE radius of the primary, and $\alpha_{\textsc{ce},\textrm{min}}$ will be smaller by a factor of $\sim5$.

\subsection{Deficits of the CE scenario}

So far the CE scenario seems successful, since the system can eject the envelope with an efficiency smaller than unity $\alpha_\textsc{ce}<1$ if the companion was larger than $\sim6\msun$. However, this scenario has a difficulty in explaining the post-CE evolution of the binary. Since the final radius of the progenitor should be larger than $\sim30\rsun$ (see Fig.\ref{allowed} and \ref{allowed2}), which is much larger than the values of $a_f$ calculated above, it is almost impossible to avoid a second CE phase. The outcome of a CE phase with a naked helium star is not known. But if the binary can successfully eject the envelope again\footnote{This may lead to the ejection of the whole helium envelope, which may change the spectral type of SN to type Ic.}, it will surely shrink the orbit even more. The primary will not be able to re-expand to $\sim30\rsun$ this way. Therefore the second CE phase should have failed and the secondary star will have been engulfed by the primary before SN explosion. If the secondary was a main sequence star, there will be a substantial amount of fresh hydrogen injected to the core of the primary. This can significantly alter the appearance of the progenitor, taking it away from the allowed region and also may change the spectral type of the SN. The mass of the secondary should also be very small in order to keep the ejecta mass smaller than $\lesssim2\msun$. But the first CE phase will not have succeeded in the first place if the mass was so small, unless the $\alpha_\textsc{ce}$ parameter is considerably large. Therefore the secondary should avoid the second CE phase or be completely lost before SN. In order to avoid the second CE phase, the post-CE separation should be large enough so that the primary never interacts with the secondary again. In the upper panel of Fig.\ref{fig:minalpha}, we show the minimum $\alpha_\textsc{ce}$ required to have a large enough post-CE separation so that the Roche lobe radius for the primary becomes $30\rsun$. The required value for $\alpha_\textsc{ce}$ becomes $\gtrsim6$ even for the largest possible secondary masses, which is very unlikely even with the consideration of other energy sources such as recombination energy. The primary may have lost its companion because of a third body encounter, but this may also be difficult considering the very tight post-CE orbit. Unless we resolve this problem, the CE scenario should be refuted.

To sum up, the CE scenario is able to reproduce the observed ejecta mass, pre- and post-SN photometry. However, the success of this model requires a significant orbital shrinkage, which will suffer a second CE phase before SN. The second CE phase will ruin the advantages of this model by increasing the ejecta mass and altering the pre-SN photometry. The ejected oxygen mass may also be smaller than the observed amount. Therefore we conclude here that the CE scenario is not suitable to explain the formation of the progenitor of iPTF13bvn.

\section{Stable Mass Transfer to a Black Hole?}
From the previous discussion, the CE scenario seems not to be suitable as the formation scenario of the progenitor of iPTF13bvn. Here we will return to the stable mass transfer scenario again. The reason that we have excluded this scenario in the first place was the non-detection of a companion. A sufficiently large companion is needed to enable stable mass transfer, and the star will also grow due to the accretion of transferred mass. However, this is only problematic if the companion is on the main sequence. The situation will be completely different if the secondary was a BH, since we can not observe a BH whatever the mass is unless it has an accretion disc around it.

\begin{figure}
 \includegraphics[width=\linewidth]{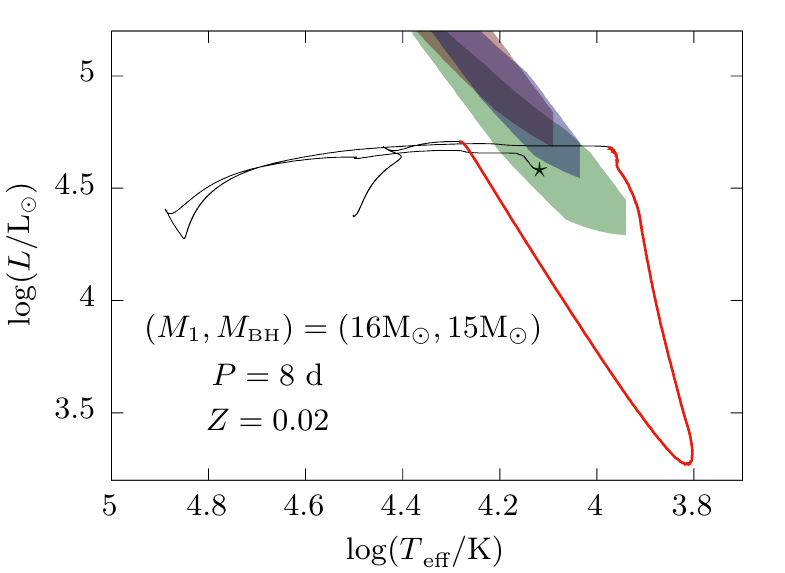}
 \caption{Evolutionary track of a $16\msun$ star with a $15\msun$ BH companion. The star symbol marks the position eight years before SN. The red part of the curve indicates the mass transfer phase.\label{BHbinaryHRD}}
\end{figure}

Here we will demonstrate that a binary with a BH component can evolve up to SN without conflicting with any of the observational constraints. We used the binary module in \texttt{MESA}, and simulated the evolution of a $16\msun$ star with a $15\msun$ BH companion in an 8 day circular orbit. The metallicity is assumed to be $Z=0.02$. The mass transfer rate was calculated according to the prescription by \citet{kol90} and the mass retention on the BH was limited by the Eddington limit. The evolutionary track of the primary is shown in Fig.\ref{BHbinaryHRD}, overplotted on the allowed region again. This system undergoes a case B mass tranfer, losing most of its hydrogen envelope during this phase. When the remaining hydrogen becomes sufficiently small, the star contracts rapidly and detaches from the BH. Most of the remaining hydrogen is burned in the H burning shell and only $\lesssim0.04\msun$ is left by the time of SN. This small amount of hydrogen may be the origin of the weak H$\alpha$ lines in the early spectra \citep[]{fre16}. The endpoint of the evolution rests in the allowed region, which makes this system a good candidate for the progenitor of iPTF13bvn. There is almost no change in the mass of the BH, only growing by $\sim0.017\msun$, meaning that most of the mass has been lost from the system. The overall evolution of the primary does not change largely even if we increase the mass of the BH up to $\sim100\msun$.

This demonstration is only an example of the evolutionary path, and there is a wider range of possible initial parameters. The primary mass should be in the range $\sim14$--17$\msun$ to create a He core in the mass range of the observed ejecta mass, create the right amount of oxygen, and have a final luminosity consistent with the pre-SN fluxes. On the other hand, there is no strong constraint on the BH mass because it does not largely affect the evolution of the primary. The only constraint is the lower limit which is roughly $\sim0.8$ times the primary mass, to enable stable mass transfer. The initial period range should be roughly 4--20 days for the mass transfer to initiate in case B, although we can not rule out case A mass transferring models\footnote{The mass range discussed here are quite sensitive to the parameters and assumptions applied to the stellar evolution code such as the mixing length, overshoot parameters or convection criteria.}.

The largest uncertainty in this model is the origin of the BH. For example, the BH could have been an extremely massive star ($\gtrsim30\msun$), in a relatively wide orbit with a $\sim14$--$17\msun$ companion. As the more massive star evolves, it develops a $\gtrsim15\msun$ He core, and expands to $\sim2000\rsun$. This may initiate a CE episode, and because the core mass is large, the post-CE separation is at moderate distances $\sim50\rsun$. At some point the massive He core will collapse to a BH and then the system will follow an evolution similar to that in the above demonstration. There is no strong support to this scenario, and deeper investigations should be carried out to check the relevance of this model. We will leave this to future works.

Since the ejecta mass is much smaller than the expected BH mass, the system will still be bound after the SN explosion, and the outcome of this model will be a BH-NS binary in a relatively wide orbit ($\sim100\rsun$). Thus we expect that we will not find a companion star in any future observations. The orbital separation is too wide to cause a BH-NS merger in a realistic time-scale, leaving no hope for gravitational wave detection. It will be extremely difficult to confirm our scenario, but the non-detection of a companion star in the next few years can strengthen our hypothesis.

\section{Conclusions}\label{sec:conclusion}
The observational constraints on the progenitor of iPTF13bvn have been revisited. We evaluated the possible position on the HR diagram and constrained the photospheric radius of the progenitor. The radius should have been in the range $\sim$20--70$\rsun$. All studies now agree that the progenitor should have been in a binary system, and expect to be able to detect a companion star in the future. We have derived the upper limit on the remaining secondary star based on the latest observational data of the SN and obtained similar results to previous works ($\sim20\msun$). But this is probably much smaller if we consider the effects of SN ejecta-companion interaction as discussed in \citet{RH15}.

We have also reassessed the relevance of the formation scenario of the progenitor via a CE phase. We performed stellar evolution calculations to mimick the post-CE evolutionary tracks just for the donor, and found a model that matches the observational constraints. However, if we consider the energy budget in the CE phase, an extremely large companion or a very high CE efficiency ($\gtrsim6$) is required to avoid a second CE phase, which is unrealistic. Therefore we conclude that the CE scenario is unlikely to be the formation scenario for the progenitor of iPTF13bvn.

As an alternative model, we considered the evolution of a binary with a BH component. Stable mass transfer from the primary star to a BH can strip off most of the hydrogen envelope up to the edge of the He core. We have demonstrated one example evolutionary track that satisfies all observational constraints. We roughly estimate that the mass of the primary should be in the range $\sim14$--$17\msun$ and the BH should be heavier than 0.8 times the primary mass. The orbital period should be $\sim$4--20 days.

There is still no quantitative support on the origin of the BH. The system may have experienced a CE phase of a much larger star, but it remains a matter of speculation. We will leave the investigation to future works.

It is almost impossible to confirm our scenario by future observations because the expected outcome is a wide BH-NS binary. However, if there is no detection of a companion star in the coming few years, we believe it will be a strong support of our model.

\section*{Acknowledgements}

The author would like to thank Shoichi Yamada, Yu Yamamoto, Keiichi Maeda, Yuichiro Sekiguchi and Pablo Marchant for useful discussions. This research has made use of the NASA/IPAC Extragalactic Database (NED) which is operated by the Jet Propulsion Laboratory, California Institute of Technology, under contract with the National Aeronautics and Space Administration. This work was supported by JSPS Research fellowship for young scientists (DC2, No 16J07613).

\bibliographystyle{mnras}

\begin{thebibliography}{}
\makeatletter
\relax
\def\mn@urlcharsother{\let\do\@makeother \do\$\do\&\do\#\do\^\do\_\do\%\do\~}
\def\mn@doi{\begingroup\mn@urlcharsother \@ifnextchar [ {\mn@doi@}
  {\mn@doi@[]}}
\def\mn@doi@[#1]#2{\def\@tempa{#1}\ifx\@tempa\@empty \href
  {http://dx.doi.org/#2} {doi:#2}\else \href {http://dx.doi.org/#2} {#1}\fi
  \endgroup}
\def\mn@eprint#1#2{\mn@eprint@#1:#2::\@nil}
\def\mn@eprint@arXiv#1{\href {http://arxiv.org/abs/#1} {{\tt arXiv:#1}}}
\def\mn@eprint@dblp#1{\href {http://dblp.uni-trier.de/rec/bibtex/#1.xml}
  {dblp:#1}}
\def\mn@eprint@#1:#2:#3:#4\@nil{\def\@tempa {#1}\def\@tempb {#2}\def\@tempc
  {#3}\ifx \@tempc \@empty \let \@tempc \@tempb \let \@tempb \@tempa \fi \ifx
  \@tempb \@empty \def\@tempb {arXiv}\fi \@ifundefined
  {mn@eprint@\@tempb}{\@tempb:\@tempc}{\expandafter \expandafter \csname
  mn@eprint@\@tempb\endcsname \expandafter{\@tempc}}}

\bibitem[\protect\citeauthoryear{{Belczynski}, {Kalogera}  \&
  {Bulik}}{{Belczynski} et~al.}{2002}]{bel02}
{Belczynski} K.,  {Kalogera} V.,   {Bulik} T.,  2002, \mn@doi [\apj]
  {10.1086/340304}, \href {http://adsabs.harvard.edu/abs/2002ApJ...572..407B}
  {572, 407}

\bibitem[\protect\citeauthoryear{{Bersten} et~al.,}{{Bersten}
  et~al.}{2014}]{ber14}
{Bersten} M.~C.,  et~al., 2014, \mn@doi [\aj] {10.1088/0004-6256/148/4/68},
  \href {http://adsabs.harvard.edu/abs/2014AJ....148...68B} {148, 68}

\bibitem[\protect\citeauthoryear{Blagorodnova et~al.,}{Blagorodnova
  et~al.}{2017}]{bla16}
Blagorodnova N.,  et~al., 2017, \apj, 834, 107

\bibitem[\protect\citeauthoryear{{Cao} et~al.,}{{Cao} et~al.}{2013}]{cao13}
{Cao} Y.,  et~al., 2013, \mn@doi [\apjl] {10.1088/2041-8205/775/1/L7}, \href
  {http://adsabs.harvard.edu/abs/2013ApJ...775L...7C} {775, L7}

\bibitem[\protect\citeauthoryear{{Cardelli}, {Clayton}  \& {Mathis}}{{Cardelli}
  et~al.}{1989}]{car89}
{Cardelli} J.~A.,  {Clayton} G.~C.,   {Mathis} J.~S.,  1989, \mn@doi [\apj]
  {10.1086/167900}, \href {http://adsabs.harvard.edu/abs/1989ApJ...345..245C}
  {345, 245}

\bibitem[\protect\citeauthoryear{{Chesneau} et~al.,}{{Chesneau}
  et~al.}{2014}]{che14}
{Chesneau} O.,  et~al., 2014, \mn@doi [\aap] {10.1051/0004-6361/201424458},
  \href {http://adsabs.harvard.edu/abs/2014A%26A...569L...3C} {569, L3}

\bibitem[\protect\citeauthoryear{{Eggleton}}{{Eggleton}}{1983}]{egg83}
{Eggleton} P.~P.,  1983, \mn@doi [\apj] {10.1086/160960}, \href
  {http://adsabs.harvard.edu/abs/1983ApJ...268..368E} {268, 368}

\bibitem[\protect\citeauthoryear{{Eggleton}}{{Eggleton}}{2011}]{egg11}
{Eggleton} P.,  2011, {Evolutionary Processes in Binary and Multiple Stars}

\bibitem[\protect\citeauthoryear{{Eldridge} \& {Maund}}{{Eldridge} \&
  {Maund}}{2016}]{eld16}
{Eldridge} J.~J.,  {Maund} J.~R.,  2016, \mn@doi [\mnras]
  {10.1093/mnrasl/slw099}, \href
  {http://adsabs.harvard.edu/abs/2016MNRAS.461L.117E} {461, L117}

\bibitem[\protect\citeauthoryear{{Eldridge}, {Izzard}  \& {Tout}}{{Eldridge}
  et~al.}{2008}]{eld08}
{Eldridge} J.~J.,  {Izzard} R.~G.,   {Tout} C.~A.,  2008, \mn@doi [\mnras]
  {10.1111/j.1365-2966.2007.12738.x}, \href
  {http://adsabs.harvard.edu/abs/2008MNRAS.384.1109E} {384, 1109}

\bibitem[\protect\citeauthoryear{{Eldridge}, {Fraser}, {Maund}  \&
  {Smartt}}{{Eldridge} et~al.}{2015}]{eld15}
{Eldridge} J.~J.,  {Fraser} M.,  {Maund} J.~R.,   {Smartt} S.~J.,  2015,
  \mn@doi [\mnras] {10.1093/mnras/stu2197}, \href
  {http://adsabs.harvard.edu/abs/2015MNRAS.446.2689E} {446, 2689}

\bibitem[\protect\citeauthoryear{{Folatelli} et~al.,}{{Folatelli}
  et~al.}{2016}]{fol16}
{Folatelli} G.,  et~al., 2016, \mn@doi [\apjl] {10.3847/2041-8205/825/2/L22},
  \href {http://adsabs.harvard.edu/abs/2016ApJ...825L..22F} {825, L22}

\bibitem[\protect\citeauthoryear{{Fremling} et~al.,}{{Fremling}
  et~al.}{2014}]{fre14}
{Fremling} C.,  et~al., 2014, \mn@doi [\aap] {10.1051/0004-6361/201423884},
  \href {http://adsabs.harvard.edu/abs/2014A%26A...565A.114F} {565, A114}

\bibitem[\protect\citeauthoryear{{Fremling} et~al.,}{{Fremling}
  et~al.}{2016}]{fre16}
{Fremling} C.,  et~al., 2016, \mn@doi [\aap] {10.1051/0004-6361/201628275},
  \href {http://adsabs.harvard.edu/abs/2016A%26A...593A..68F} {593, A68}

\bibitem[\protect\citeauthoryear{{Ge}, {Hjellming}, {Webbink}, {Chen}  \&
  {Han}}{{Ge} et~al.}{2010}]{ge10}
{Ge} H.,  {Hjellming} M.~S.,  {Webbink} R.~F.,  {Chen} X.,   {Han} Z.,  2010,
  \mn@doi [\apj] {10.1088/0004-637X/717/2/724}, \href
  {http://adsabs.harvard.edu/abs/2010ApJ...717..724G} {717, 724}

\bibitem[\protect\citeauthoryear{{Groh}, {Georgy}  \& {Ekstr{\"o}m}}{{Groh}
  et~al.}{2013}]{gro13}
{Groh} J.~H.,  {Georgy} C.,   {Ekstr{\"o}m} S.,  2013, \mn@doi [\aap]
  {10.1051/0004-6361/201322369}, \href
  {http://adsabs.harvard.edu/abs/2013A%26A...558L...1G} {558, L1}

\bibitem[\protect\citeauthoryear{{Hall} \& {Tout}}{{Hall} \&
  {Tout}}{2014}]{hal14}
{Hall} P.~D.,  {Tout} C.~A.,  2014, \mn@doi [\mnras] {10.1093/mnras/stu1678},
  \href {http://adsabs.harvard.edu/abs/2014MNRAS.444.3209H} {444, 3209}

\bibitem[\protect\citeauthoryear{{Hirai} \& {Yamada}}{{Hirai} \&
  {Yamada}}{2015}]{RH15}
{Hirai} R.,  {Yamada} S.,  2015, \mn@doi [\apj] {10.1088/0004-637X/805/2/170},
  \href {http://adsabs.harvard.edu/abs/2015ApJ...805..170H} {805, 170}

\bibitem[\protect\citeauthoryear{{Hut}}{{Hut}}{1980}]{hut80}
{Hut} P.,  1980, \aap, \href
  {http://adsabs.harvard.edu/abs/1980A%26A....92..167H} {92, 167}

\bibitem[\protect\citeauthoryear{{Iaconi}, {Reichardt}, {Staff}, {De Marco},
  {Passy}, {Price}, {Wurster}  \& {Herwig}}{{Iaconi} et~al.}{2016}]{iac16}
{Iaconi} R.,  {Reichardt} T.,  {Staff} J.,  {De Marco} O.,  {Passy} J.-C.,
  {Price} D.,  {Wurster} J.,   {Herwig} F.,  2016, \mn@doi [\mnras]
  {10.1093/mnras/stw2377}, \href
  {http://adsabs.harvard.edu/abs/2016MNRAS.tmp.1479I} {}

\bibitem[\protect\citeauthoryear{{Iben} \& {Tutukov}}{{Iben} \&
  {Tutukov}}{1984}]{ibe84}
{Iben} Jr. I.,  {Tutukov} A.~V.,  1984, \mn@doi [\apj] {10.1086/162455}, \href
  {http://adsabs.harvard.edu/abs/1984ApJ...284..719I} {284, 719}

\bibitem[\protect\citeauthoryear{{Ivanova}}{{Ivanova}}{2011}]{iva11}
{Ivanova} N.,  2011, \mn@doi [\apj] {10.1088/0004-637X/730/2/76}, \href
  {http://adsabs.harvard.edu/abs/2011ApJ...730...76I} {730, 76}

\bibitem[\protect\citeauthoryear{{Ivanova} et~al.,}{{Ivanova}
  et~al.}{2013a}]{iva13}
{Ivanova} N.,  et~al., 2013a, \mn@doi [\aapr] {10.1007/s00159-013-0059-2},
  \href {http://adsabs.harvard.edu/abs/2013A%26ARv..21...59I} {21, 59}

\bibitem[\protect\citeauthoryear{{Ivanova}, {Justham}, {Avendano Nandez}  \&
  {Lombardi}}{{Ivanova} et~al.}{2013b}]{iva13b}
{Ivanova} N.,  {Justham} S.,  {Avendano Nandez} J.~L.,   {Lombardi} J.~C.,
  2013b, \mn@doi [Science] {10.1126/science.1225540}, \href
  {http://adsabs.harvard.edu/abs/2013Sci...339..433I} {339, 433}

\bibitem[\protect\citeauthoryear{{Jerkstrand}, {Ergon}, {Smartt}, {Fransson},
  {Sollerman}, {Taubenberger}, {Bersten}  \& {Spyromilio}}{{Jerkstrand}
  et~al.}{2015}]{jer15}
{Jerkstrand} A.,  {Ergon} M.,  {Smartt} S.~J.,  {Fransson} C.,  {Sollerman} J.,
   {Taubenberger} S.,  {Bersten} M.,   {Spyromilio} J.,  2015, \mn@doi [\aap]
  {10.1051/0004-6361/201423983}, \href
  {http://adsabs.harvard.edu/abs/2015A%26A...573A..12J} {573, A12}

\bibitem[\protect\citeauthoryear{{Kalogera}, {Belczynski}, {Kim},
  {O'Shaughnessy}  \& {Willems}}{{Kalogera} et~al.}{2007}]{kal07}
{Kalogera} V.,  {Belczynski} K.,  {Kim} C.,  {O'Shaughnessy} R.,   {Willems}
  B.,  2007, \mn@doi [\physrep] {10.1016/j.physrep.2007.02.008}, \href
  {http://adsabs.harvard.edu/abs/2007PhR...442...75K} {442, 75}

\bibitem[\protect\citeauthoryear{{Kolb} \& {Ritter}}{{Kolb} \&
  {Ritter}}{1990}]{kol90}
{Kolb} U.,  {Ritter} H.,  1990, \aap, \href
  {http://adsabs.harvard.edu/abs/1990A%26A...236..385K} {236, 385}

\bibitem[\protect\citeauthoryear{{Kuncarayakti} et~al.,}{{Kuncarayakti}
  et~al.}{2015}]{kun15}
{Kuncarayakti} H.,  et~al., 2015, \mn@doi [\aap] {10.1051/0004-6361/201425604},
  \href {http://adsabs.harvard.edu/abs/2015A%26A...579A..95K} {579, A95}

\bibitem[\protect\citeauthoryear{{Lai}, {Rasio}  \& {Shapiro}}{{Lai}
  et~al.}{1993}]{lai93}
{Lai} D.,  {Rasio} F.~A.,   {Shapiro} S.~L.,  1993, \mn@doi [\apjl]
  {10.1086/186787}, \href {http://adsabs.harvard.edu/abs/1993ApJ...406L..63L}
  {406, L63}

\bibitem[\protect\citeauthoryear{{Limongi} \& {Chieffi}}{{Limongi} \&
  {Chieffi}}{2003}]{lim03}
{Limongi} M.,  {Chieffi} A.,  2003, \mn@doi [\apj] {10.1086/375703}, \href
  {http://adsabs.harvard.edu/abs/2003ApJ...592..404L} {592, 404}

\bibitem[\protect\citeauthoryear{{MacLeod} \& {Ramirez-Ruiz}}{{MacLeod} \&
  {Ramirez-Ruiz}}{2015}]{mac15}
{MacLeod} M.,  {Ramirez-Ruiz} E.,  2015, \mn@doi [\apjl]
  {10.1088/2041-8205/798/1/L19}, \href
  {http://adsabs.harvard.edu/abs/2015ApJ...798L..19M} {798, L19}

\bibitem[\protect\citeauthoryear{{MacLeod}, {Macias}, {Ramirez-Ruiz},
  {Grindlay}, {Batta}  \& {Montes}}{{MacLeod} et~al.}{2016}]{mac15b}
{MacLeod} M.,  {Macias} P.,  {Ramirez-Ruiz} E.,  {Grindlay} J.,  {Batta} A.,
  {Montes} G.,  2016, preprint, \href
  {http://adsabs.harvard.edu/abs/2016arXiv160501493M} {} (\mn@eprint {arXiv}
  {1605.01493})

\bibitem[\protect\citeauthoryear{{Maund}, {Reilly}  \& {Mattila}}{{Maund}
  et~al.}{2014}]{mau14}
{Maund} J.~R.,  {Reilly} E.,   {Mattila} S.,  2014, \mn@doi [\mnras]
  {10.1093/mnras/stt2131}, \href
  {http://adsabs.harvard.edu/abs/2014MNRAS.438..938M} {438, 938}

\bibitem[\protect\citeauthoryear{{Nandez}, {Ivanova}  \& {Lombardi}}{{Nandez}
  et~al.}{2014}]{nan14}
{Nandez} J.~L.~A.,  {Ivanova} N.,   {Lombardi} Jr. J.~C.,  2014, \mn@doi [\apj]
  {10.1088/0004-637X/786/1/39}, \href
  {http://adsabs.harvard.edu/abs/2014ApJ...786...39N} {786, 39}

\bibitem[\protect\citeauthoryear{{Nomoto}, {Hashimoto}, {Tsujimoto},
  {Thielemann}, {Kishimoto}, {Kubo}  \& {Nakasato}}{{Nomoto}
  et~al.}{1997}]{nom97}
{Nomoto} K.,  {Hashimoto} M.,  {Tsujimoto} T.,  {Thielemann} F.-K.,
  {Kishimoto} N.,  {Kubo} Y.,   {Nakasato} N.,  1997, \mn@doi [Nuclear Physics
  A] {10.1016/S0375-9474(97)00076-6}, \href
  {http://adsabs.harvard.edu/abs/1997NuPhA.616...79N} {616, 79}

\bibitem[\protect\citeauthoryear{{Ohlmann}, {R{\"o}pke}, {Pakmor}, {Springel}
  \& {M{\"u}ller}}{{Ohlmann} et~al.}{2016a}]{ohl16a}
{Ohlmann} S.~T.,  {R{\"o}pke} F.~K.,  {Pakmor} R.,  {Springel} V.,
  {M{\"u}ller} E.,  2016a, \mn@doi [\mnras] {10.1093/mnrasl/slw144}, \href
  {http://adsabs.harvard.edu/abs/2016MNRAS.462L.121O} {462, L121}

\bibitem[\protect\citeauthoryear{{Ohlmann}, {R{\"o}pke}, {Pakmor}  \&
  {Springel}}{{Ohlmann} et~al.}{2016b}]{ohl16b}
{Ohlmann} S.~T.,  {R{\"o}pke} F.~K.,  {Pakmor} R.,   {Springel} V.,  2016b,
  \mn@doi [\apjl] {10.3847/2041-8205/816/1/L9}, \href
  {http://adsabs.harvard.edu/abs/2016ApJ...816L...9O} {816, L9}

\bibitem[\protect\citeauthoryear{{Passy} et~al.,}{{Passy} et~al.}{2012}]{pas12}
{Passy} J.-C.,  et~al., 2012, \mn@doi [\apj] {10.1088/0004-637X/744/1/52},
  \href {http://adsabs.harvard.edu/abs/2012ApJ...744...52P} {744, 52}

\bibitem[\protect\citeauthoryear{{Paxton}, {Bildsten}, {Dotter}, {Herwig},
  {Lesaffre}  \& {Timmes}}{{Paxton} et~al.}{2011}]{MESA1}
{Paxton} B.,  {Bildsten} L.,  {Dotter} A.,  {Herwig} F.,  {Lesaffre} P.,
  {Timmes} F.,  2011, \mn@doi [\apjs] {10.1088/0067-0049/192/1/3}, \href
  {http://adsabs.harvard.edu/abs/2011ApJS..192....3P} {192, 3}

\bibitem[\protect\citeauthoryear{{Paxton} et~al.,}{{Paxton}
  et~al.}{2013}]{MESA2}
{Paxton} B.,  et~al., 2013, \mn@doi [\apjs] {10.1088/0067-0049/208/1/4}, \href
  {http://adsabs.harvard.edu/abs/2013ApJS..208....4P} {208, 4}

\bibitem[\protect\citeauthoryear{{Paxton} et~al.,}{{Paxton}
  et~al.}{2015}]{MESA3}
{Paxton} B.,  et~al., 2015, \mn@doi [\apjs] {10.1088/0067-0049/220/1/15}, \href
  {http://adsabs.harvard.edu/abs/2015ApJS..220...15P} {220, 15}

\bibitem[\protect\citeauthoryear{{Rauscher}, {Heger}, {Hoffman}  \&
  {Woosley}}{{Rauscher} et~al.}{2002}]{rau02}
{Rauscher} T.,  {Heger} A.,  {Hoffman} R.~D.,   {Woosley} S.~E.,  2002, \mn@doi
  [\apj] {10.1086/341728}, \href
  {http://adsabs.harvard.edu/abs/2002ApJ...576..323R} {576, 323}

\bibitem[\protect\citeauthoryear{{Ricker} \& {Taam}}{{Ricker} \&
  {Taam}}{2012}]{ric12}
{Ricker} P.~M.,  {Taam} R.~E.,  2012, \mn@doi [\apj]
  {10.1088/0004-637X/746/1/74}, \href
  {http://adsabs.harvard.edu/abs/2012ApJ...746...74R} {746, 74}

\bibitem[\protect\citeauthoryear{{Schlafly} \& {Finkbeiner}}{{Schlafly} \&
  {Finkbeiner}}{2011}]{sch11}
{Schlafly} E.~F.,  {Finkbeiner} D.~P.,  2011, \mn@doi [\apj]
  {10.1088/0004-637X/737/2/103}, \href
  {http://adsabs.harvard.edu/abs/2011ApJ...737..103S} {737, 103}

\bibitem[\protect\citeauthoryear{{Smith} \& {Arnett}}{{Smith} \&
  {Arnett}}{2014}]{smi14}
{Smith} N.,  {Arnett} W.~D.,  2014, \mn@doi [\apj]
  {10.1088/0004-637X/785/2/82}, \href
  {http://adsabs.harvard.edu/abs/2014ApJ...785...82S} {785, 82}

\bibitem[\protect\citeauthoryear{{Srivastav}, {Anupama}  \& {Sahu}}{{Srivastav}
  et~al.}{2014}]{sri14}
{Srivastav} S.,  {Anupama} G.~C.,   {Sahu} D.~K.,  2014, \mn@doi [\mnras]
  {10.1093/mnras/stu1878}, \href
  {http://adsabs.harvard.edu/abs/2014MNRAS.445.1932S} {445, 1932}

\bibitem[\protect\citeauthoryear{{Tully}, {Rizzi}, {Shaya}, {Courtois},
  {Makarov}  \& {Jacobs}}{{Tully} et~al.}{2009}]{tul09}
{Tully} R.~B.,  {Rizzi} L.,  {Shaya} E.~J.,  {Courtois} H.~M.,  {Makarov}
  D.~I.,   {Jacobs} B.~A.,  2009, \mn@doi [\aj] {10.1088/0004-6256/138/2/323},
  \href {http://adsabs.harvard.edu/abs/2009AJ....138..323T} {138, 323}

\bibitem[\protect\citeauthoryear{{Tully} et~al.,}{{Tully} et~al.}{2013}]{tul13}
{Tully} R.~B.,  et~al., 2013, \mn@doi [\aj] {10.1088/0004-6256/146/4/86}, \href
  {http://adsabs.harvard.edu/abs/2013AJ....146...86T} {146, 86}

\bibitem[\protect\citeauthoryear{{Wang}, {Jia}  \& {Li}}{{Wang}
  et~al.}{2016}]{wan16}
{Wang} C.,  {Jia} K.,   {Li} X.-D.,  2016, \mn@doi [Research in Astronomy and
  Astrophysics] {10.1088/1674-4527/16/8/126}, \href
  {http://adsabs.harvard.edu/abs/2016RAA....16..126W} {16, 126}

\bibitem[\protect\citeauthoryear{{Webbink}}{{Webbink}}{1984}]{web84}
{Webbink} R.~F.,  1984, \mn@doi [\apj] {10.1086/161701}, \href
  {http://adsabs.harvard.edu/abs/1984ApJ...277..355W} {277, 355}

\bibitem[\protect\citeauthoryear{{Woosley} \& {Heger}}{{Woosley} \&
  {Heger}}{2007}]{woo07}
{Woosley} S.~E.,  {Heger} A.,  2007, \mn@doi [\physrep]
  {10.1016/j.physrep.2007.02.009}, \href
  {http://adsabs.harvard.edu/abs/2007PhR...442..269W} {442, 269}

\bibitem[\protect\citeauthoryear{{Xu} \& {Li}}{{Xu} \& {Li}}{2010}]{xu10}
{Xu} X.-J.,  {Li} X.-D.,  2010, \mn@doi [\apj] {10.1088/0004-637X/716/1/114},
  \href {http://adsabs.harvard.edu/abs/2010ApJ...716..114X} {716, 114}

\bibitem[\protect\citeauthoryear{{Yoon}, {Gr{\"a}fener}, {Vink}, {Kozyreva}  \&
  {Izzard}}{{Yoon} et~al.}{2012}]{yoo12}
{Yoon} S.-C.,  {Gr{\"a}fener} G.,  {Vink} J.~S.,  {Kozyreva} A.,   {Izzard}
  R.~G.,  2012, \mn@doi [\aap] {10.1051/0004-6361/201219790}, \href
  {http://adsabs.harvard.edu/abs/2012A%26A...544L..11Y} {544, L11}

\bibitem[\protect\citeauthoryear{{de Jager}, {Nieuwenhuijzen}  \& {van der
  Hucht}}{{de Jager} et~al.}{1988}]{dej88}
{de Jager} C.,  {Nieuwenhuijzen} H.,   {van der Hucht} K.~A.,  1988, \aaps,
  \href {http://adsabs.harvard.edu/abs/1988A%26AS...72..259D} {72, 259}

\bibitem[\protect\citeauthoryear{{de Kool}}{{de Kool}}{1990}]{dek90}
{de Kool} M.,  1990, \mn@doi [\apj] {10.1086/168974}, \href
  {http://adsabs.harvard.edu/abs/1990ApJ...358..189D} {358, 189}

\bibitem[\protect\citeauthoryear{{van den Heuvel}}{{van den
  Heuvel}}{1976}]{van76}
{van den Heuvel} E.~P.~J.,  1976, in {Eggleton} P.,  {Mitton} S.,   {Whelan}
  J.,  eds,  IAU Symposium Vol. 73, Structure and Evolution of Close Binary
  Systems. p.~35

\makeatother
\end{thebibliography}

\bsp
\label{lastpage}
\end{document}